\documentclass[aps,prc,reprint,showpacs]{revtex4-1}
\usepackage{graphicx}
\usepackage{hyperref}
\usepackage{subfigure}

\newcommand{ \be }{\begin{equation}}
\newcommand{ \ee }{\end{equation}}
\newcommand{ \bea }{\begin{eqnarray}}
\newcommand{ \eea }{\end{eqnarray}}
\newcommand{ \la }{\langle}
\newcommand{ \ra }{\rangle}
\newcommand{ \lp }{\left(}
\newcommand{ \rp }{\right)}

\newcommand{ \R }{{\cal R}}

\begin{document}

\title{Fluctuation Probes of Early-Time Correlations in Nuclear Collisions}
%\title{Fluctuation and Correlation Probes of Early-Time Dynamics in Nuclear Collisions}

\author{Sean Gavin$^a$ and George Moschelli$^b$}
%\email[]{sean@physics.wayne.edu}
\affiliation{a) Department of Physics and Astronomy, Wayne State University, 666 W Hancock, Detroit,
MI, 48202, USA\\
b) Frankfurt Institute for Advanced Studies, Johann Wolfgang Goethe University,
Ruth-Moufang-Str.~1, 60438 Frankfurt am Main, Germany}

\date{\today}

\begin{abstract}
Correlation measurements imply that anisotropic flow in nuclear collisions includes a novel triangular component along with the more familiar elliptic-flow contribution. Triangular flow has been attributed to event-wise fluctuations in the initial shape of the collision volume.  We ask two questions: 1) How do these shape fluctuations impact other event-by-event observables? 2) Can we disentangle fundamental information on the early time fluctuations from the complex flow that results? We study correlation and fluctuation observables in a framework in which flux tubes in an early Glasma stage later produce hydrodynamic flow. 
%We show how flow observables $v_n$ can be combined with multiplicity and transverse momentum fluctuations to disentangle Glasma information from hydrodynamic effects. 
Calculated multiplicity and transverse momentum fluctuations are in excellent agreement with data from 62.4 GeV Au+Au up to 2.76 TeV Pb+Pb.  
\end{abstract}

%\keywords{}

\maketitle

%
%	SECTION:	INTRODUCTION
%
\section{\label{intro}Introduction}

Measurements of two particle correlations in nuclear collisions exhibit a complex pattern of ridges, bumps, and valleys as functions of relative pseudorapidity $\Delta\eta = \eta_1-\eta_2$  and azimuthal angle $\Delta\phi = \phi_1-\phi_2$ \cite{Putschke:2007mi,Abelev:2009qa,Nattrass:2008tw,Daugherity:2008su,DeSilva:2009yy,Adams:2005aw,Adamova:2008sx,Wenger:2008ts,Alver:2009id}.  In a recent advance, Alver and Roland showed that much of the $\Delta\phi$ dependence of  these correlations can be described in terms of anisotropic flow \cite{Alver:2010gr}. Their key realization was that a novel triangular flow contribution $ \sim v_3 \cos 3\Delta\phi $ is needed to describe the data. Such contributions have been attributed to fluctuations of the geometric shape of the collision volume from event to event \cite{Takahashi:2009na,Alver:2010gr,Alver:2010dn,Sorensen:2010zq,Sorensen:2011hm,Luzum:2010sp,Petersen:2010zt,Qin:2010pf,Petersen:2010md,Teaney:2010vd,Schenke:2010rr,Qiu:2011iv}.  A unique shape is determined in the first instants of each collision event as the nuclei crash through one another. 

Interestingly, fluctuations have been of measured in nuclear collisions for many years, but for different reasons.  Event-by-event fluctuations of the multiplicity, mean transverse momentum, and net charge are viewed as probes of the QCD phase transition, although signature behavior has yet to be seen \cite{Westfall:2008zz,Koch:2008ia,Aggarwal:2010cw}.  Such observations are explained by the variation of density, geometry, thermalization, and flow; see, e.g., \cite{Gavin:2003cb,Gavin:2004dc,Voloshin:2003ud,Mrowczynski:2004cg,Broniowski:2009fm}. 

In this paper we explore the common influence that the early-time dynamics of the system has on correlations, flow, and fluctuations.  We begin by asking how geometric  fluctuations impact other event-by-event observables. In Sec.\ \ref{corrFluc}, we recall the relationship between correlations and anisotropic flow. Next, we discuss the relationship between fluctuations and correlations in Sec.\ \ref{Fluc}. We exploit integral relationships obtained in Ref.\ \cite{Pruneau:2002yf} to marry the dynamical language of flow and correlations to the statistical formulation of fluctuation observables. This yields model independent properties of experimental observables. In particular, we find that geometric shape fluctuations alone cannot explain measured multiplicity and transverse momentum fluctuations.

We next ask what fluctuation and correlation measurements can reveal about the early-time dynamics of the collision system. The first evidence that correlations originate at early times in the collision is the long rapidity range of the ridge \cite{Dumitru:2008wn,Gavin:2008ev}.  Correlations show a ridge-like peak near $\Delta\phi = 0$.  This ridge extends over a broad range in $\Delta\eta$, as do away-side features centered near  $\Delta\phi \sim \pi$ \cite{Wenger:2008ts}. Causality dictates that correlations over several rapidity units must originate at the earliest stages of the collision \cite{Dumitru:2008wn,Gavin:2008ev}. With that in mind, we described the ridge as a consequence of particle production in an early Glasma stage followed by transverse flow in Refs.\  \cite{Dumitru:2008wn,Gavin:2008ev,Moschelli:2009tg,Dusling:2009ar}. Our description is part of a broader family of models in which particles are initially correlated at the point of production \cite{Voloshin:2003ud,Pruneau:2007ua,Lindenbaum:2007ui,Peitzmann:2009vj,Takahashi:2009na,Andrade:2010xy,Werner:2010aa,Sharma:2011nj}. 

Identifying the impact of anisotropic flow on these correlations adds considerable credence to this observation \cite{Sorensen:2011hm}.  Anisotropic flow is well understood as an early time effect, since it is generated in part by the geometric configuration of  participant nucleons in the colliding nuclei; see, e.g., \cite{Voloshin:2008dg,Sorensen:2009cz}. Correspondingly, the measured $v_2$ coefficients vary little with rapidity.

To illustrate how correlations, flow, and fluctuations result from the early-time dynamics, we apply the general framework of Refs.\ \cite{Gavin:2008ev,Moschelli:2009tg} in Sec.\ \ref{source}. We obtain expressions for the correlation function and its Fourier coefficients, (\ref{eq:fourier}), (\ref{eq:floCo1}), and (\ref{eq:floCo2}). This provides a unifying framework for understanding hydrodynamic $v_n$ studies together with earlier work on the ridge. We also derive expressions for transverse momentum fluctuations, (\ref{eq:dpt1}) and (\ref{eq:Gdef2}). 

In Sec.\ \ref{glasma}, we use fluctuation measurements to extract information on the particle production mechanism. We focus on the Color Glass Condensate formulation in Ref.\ \cite{Dumitru:2008wn,Gavin:2008ev,Dusling:2009ni,Moschelli:2009tg}, and argue that dynamic multiplicity fluctuations $\cal R$ can provide information on particle production that is independent of flow.  Transverse momentum fluctuations $\langle \delta p_{t1}\delta p_{t2}\rangle$  provide similar information, although the results are somewhat sensitive to radial flow (but not the $v_n$).   Constraining the flow contribution by calculating $\langle p_t\rangle$ and $v_2$,  
we compute the fluctuations for $\cal R$ and $\langle \delta p_{t1}\delta p_{t2}\rangle$. 
We find that the same model that described the energy, target-mass, and $p_t$ dependence of the ridge also describes transverse momentum fluctuations measured at the Brookhaven Relativistic Heavy Ion Collider, RHIC, and the CERN Large Hadron Collider LHC. 

%The numbers and momenta of particles differs dramatically from collision to collision due to the variation of impact parameter, energy deposition, and other dynamical effects. 

%
%	SECTION:	Correlations and Flow 
%
\section{\label{corrFluc} Correlations and Flow}

Correlation measurements commonly center on the pair distribution
\begin{equation}\label{eq:pairDensity}
    \rho_2(\mathbf{p}_1,\mathbf{p}_2) = {{dN}\over{d\eta_1d^2p_{t1}d\eta_2d^2p_{t2}}},
\end{equation}
where $\eta_i$ is pseudorapidity and $p_{ti} = |\mathbf{p}_{ti}|$ is transverse momentum of particles $i= 1,2$. In the absence of correlations, $ \rho_2(\mathbf{p}_1,\mathbf{p}_2) \rightarrow \rho_1(\mathbf{p}_1)\rho_1(\mathbf{p}_2)$, where the single particle distribution is $\rho_1(\mathbf{p}) = dN/d\eta d^2p_{t}$. Experiments typically report ratios of $\rho_2$ to either $\rho_1\rho_1$ or mixed event pairs integrated over ranges of $p_t$. To keep the notation simple, we will not make the $\eta$ dependence explicit unless needed.

To illustrate the connection between flow and correlations, recall for the moment the traditional picture of flow in which event-by-event fluctuations are neglected. It has long been known that collisions at non-zero impact parameter $b$ produce anisotropic flow \cite{Voloshin:2008dg,Sorensen:2009cz}. Anisotropy derives from the change in the shape of the collision volume with respect to the reaction plane, i.e., the plane spanned by $\mathbf{b}$ and the beam direction. The distribution with respect to this plane is 
\begin{equation}\label{eq:flow0}
\rho_1(\mathbf{p}_t; \psi_{{}_{RP}})    =  \overline{\rho}_1\{1+2\sum_{n=1}^{\infty} v_n(p_t) \cos[n(\phi-\psi_{{}_{RP}})]\},
\end{equation}
where the coefficients depend only on the magnitude $p_t= |\mathbf{p}_t|$ and $\eta$. 
An experimental analysis that does not identify the reaction plane measures the distribution $(2\pi)^{-1}\int \rho_1(\mathbf{p}_t; \psi_{{}_{RP}})  d\psi_{{}_{RP}}= \overline{\rho}_1 $, where the ``bar'' denotes average over $\phi$.  The second equality follows because $\rho_1$ is a function of $\phi-\psi_{{}_{RP}}$.  

Anisotropic flow introduces  correlations because pairs from the same collision event have the same reaction plane. The reaction plane averaged pair distribution can only be a function of the relative angle $\Delta\phi$, so that    
\begin{equation}\label{eq:flow2}
     {\rho}_2=  
         \overline{\rho}_2\{1+ 2\sum_{n=1}^{\infty} a_n(p_1,p_2) \cos(n\Delta\phi)\}. 
\end{equation}
If geometry is the only source of correlations then ${\rho}_2(\mathbf{p}_{t1}, \mathbf{p}_{t2})$ is the product $\rho_1(\mathbf{p}_{t1}; \psi_{{}_{RP}})\rho_1(\mathbf{p}_{t2}; \psi_{{}_{RP}})$ averaged over $\psi_{{}_{RP}}$. In that case the Fourier coefficients of (\ref{eq:flow0}) and (\ref{eq:flow2}) are related. For $n= 0$ we find 
\begin{equation}\label{eq:uncorrRho}
          \overline{\rho}_2(p_1, p_2)  \rightarrow  \overline{\rho}_1(p_1) \overline{\rho}_1(p_2).\,\,\,\,\,\,\,\, \,\,\,\,\,\,\,\, {\rm uncorrelated}
\end{equation}
while for $n\geq 2$ 
\begin{equation}\label{eq:uncorrA}
         a_n(p_1, p_2)  \rightarrow {v_n(p_1)v_n(p_2)}; \,\,\,\,\,\,\,\,\,\,\,\,{\rm uncorrelated}
\end{equation}
these results hold only when fluctuations may be neglected. Momentum conservation contributes to $a_1$, modifying the $v_1(p_1)v_1(p_2)$ term, as discussed by Borghini and others \cite{Borghini:2002mv,Borghini:2003ur,Borghini:2006yk,Borghini:2007ku,Chajecki:2008vg}. 

Fluctuations introduce further anisotropy because the shape of the collision volume is different in each collision event.  
In collisions of identical nuclei, the event-averaged interaction volume is symmetric in $\phi$ and fixed by $b$ and $\Psi_{{}_{RP}}$. If all events of a given $b$ had the same interaction volume, we then would expect only even $n$ to contribute to  (\ref{eq:flow0}). Shape fluctuations cause the flow parameters $v_n$ to vary from event to event and allow odd $n$ to contribute. The average pair distribution has the same form as  (\ref{eq:flow2}), but with coefficients 
\begin{equation}\label{eq:corrA}
         a_n(p_1, p_2)  = \langle {v_n(p_1)v_n(p_2)}\rangle,
\end{equation}
where the brackets denote average over events (including all event shapes) \cite{Teaney:2010vd}. The measured azimuthal dependence of two particle correlations  are reasonably described by (\ref{eq:flow2}) with $n = 1, 2$ and $3$. We emphasize that fluctuations in shape alone cannot alter $\overline{\rho}_2$ -- this requires further dynamical fluctuations which we discuss in the next sections. 

We remark in passing that one often discusses the shape fluctuations in terms of a geometric eccentricity $\epsilon_n$. If one assumes that the relation between $\epsilon_n$ and the resulting anisotropy of the fluid flow $v_n$ is approximately deterministic, then fluctuations of the ratio $v_n(p_1)/\epsilon_n$ are negligible. In that case $\langle {v_n(p_1)v_n(p_2)}\rangle\approx (v_n(p_1)/\epsilon_n)(v_n(p_2)/\epsilon_n)\langle \epsilon_n^2\rangle$. This factorization conjecture seems plausible, but currently requires further theoretical investigation  \cite{Luzum:2011mm}.

%
%	SECTION:	Fluctuations 
%
\section{\label{Fluc} Fluctuations}
Fluctuation measurements study the variation of bulk quantities, such as multiplicity or average transverse momentum, over an ensemble of events \cite{Gazdzicki:1992ri,Mrowczynski:1998vt,Westfall:2008zz}.  
Suppose that fluctuations of multiplicity $N$ result in a variance $\sigma_{{}_{N}}^2 = \langle N^2\rangle - \langle N\rangle^2$. 
Uncorrelated particles would be described by Poisson statistics,  for which $\sigma_{{}_{N}}^2\rightarrow \sigma_{\rm stat}^2 = \langle N\rangle$. Correlations give rise to a difference  $\sigma_{{}_{N}}^2  - \sigma_{\rm stat}^2$, which we characterize by
the dynamic variance  
\begin{equation}\label{eq:MultFluctExp}
    {\cal R}  ={{\langle N^2\rangle -\langle N\rangle^2 -\langle N\rangle}\over{\langle
    N\rangle^2}},
   % ={{1}\over{\langle N\rangle^2}}\int r(\mathbf{p}_{1},\mathbf{p}_{2}) d\mathbf{p}_{1}d\mathbf{p}_{2}.
\end{equation}
as discussed in~\cite{Pruneau:2002yf}.  Similarly, many describe the dynamic fluctuations of transverse momentum using the covariance 
 \begin{equation}\label{eq:ptFluctExp}
 \langle \delta p_{t1}\delta p_{t2}\rangle = 
{ {\langle \sum_{i \neq j}\delta p_{ti}\delta p_{tj}\rangle}\over{\langle N(N-1)\rangle}},
 \end{equation}
where $\delta p_{ti} = p_{ti}-\langle p_t\rangle$ and the average transverse momentum is $\langle p_t\rangle =  \langle P_t \rangle /\langle N\rangle$ for $P_t =  \sum_i p_{ti}$  the total momentum in an event \cite{Voloshin:1999yf,Voloshin:2001ei,Adamova:2003pz,Adams:2005ka}. This quantity vanishes when particles $i$ and $j$ are uncorrelated.  
Note that one can write $\langle \delta p_{t1}\delta p_{t2}\rangle$ in terms of  the variance  
$\sigma_{P_t}^2 = \langle (P_t  -N\langle p_t\rangle)^2\rangle$.  In the absence of correlations, that variance is $\sigma_{P_t\, {\rm stat}}^2 = \langle N\rangle (\langle p_t^2\rangle - \langle p_t\rangle^2)$. One can show that $ \langle \delta p_{t1}\delta p_{t2}\rangle$ is the difference $\sigma_{P_t}^2 - \sigma_{P_t\, {\rm stat}}^2$ divided by the average number of pairs $\langle N(N-1)\rangle$. 

To see the connection between fluctuation and correlation measurements, observe that the integral of the pair distribution $\rho_2$ gives the average number of pairs $\langle N(N-1)\rangle$. We then write (\ref{eq:MultFluctExp}) as
\begin{equation}\label{eq:DynamicMultDensity}
    {\cal R}  
    ={{1}\over{\langle N\rangle^2}}\int r(\mathbf{p}_{1},\mathbf{p}_{2}) d\mathbf{p}_{1}d\mathbf{p}_{2},
\end{equation}
where we define 
\begin{equation}\label{eq:corrFunExp}
    r(\mathbf{p}_1, \mathbf{p}_2) = {\rho}_2(\mathbf{p}_1,\mathbf{p}_2)
    - {\rho}_1(\mathbf{p}_1){\rho}_1(\mathbf{p}_2).  
\end{equation}
Similarly, we can write (\ref{eq:ptFluctExp}) in terms of the correlation function (\ref{eq:corrFunExp})  as 
\begin{equation}\label{eq:Dynamic}
    \langle \delta p_{t1}\delta p_{t2}\rangle =
    \int\! d\mathbf{p}_{1}d\mathbf{p}_{2}\,
    {{r(\mathbf{p}_{1},\mathbf{p}_{2})}\over{\langle N(N-1)\rangle}}
    \delta p_{t1} \delta p_{t2}.
\end{equation}
We stress that the densities $\rho_1$ and $\rho_2$ are event-averaged quantities. 

Fluctuation measurements probe the overall strength of correlations in a manner that is independent of the anisotropic flow. To see this, we combine (\ref{eq:flow2}) with (\ref{eq:corrFunExp}) to find that the $\cos n\Delta\phi$ contributions vanish on integration over $0\leq \Delta\phi\leq 2\pi$. We obtain 
\begin{equation}\label{eq:MultAns}
   {\cal R}  
    ={{1}\over{\langle N\rangle^2}}\int 
     \overline{r}(p_{t1},p_{t2})   d\mathbf{p}_{1}d\mathbf{p}_{2},
\end{equation}
which depends only on the $\phi$ averaged function
\begin{equation}\label{eq:corrFun0}
    \overline{r}(p_{t1}, {p}_{t2}) =  \overline{\rho}_2(p_{t1},p_{t2})-    \overline{\rho}_1(p_{t1})   \overline{\rho}_1(p_{t2}).  
\end{equation}
It is easy to understand why ${\cal R}$ is not sensitive anisotropy -- $N$ simply counts particles irrespective of where they are flowing. Similarly, the 
\begin{equation}\label{eq:PtAns}
    \langle \delta p_{t1}\delta p_{t2}\rangle =
    \int\! d\mathbf{p}_{1}d\mathbf{p}_{2}\,
    {{\overline{r}(p_{t1},p_{t2})}\over{\langle N(N-1)\rangle}}
    \delta p_{t1} \delta p_{t2},
\end{equation}
Here too the $\cos n\Delta\phi$ contributions to (\ref{eq:flow2}) vanish on integration. These fluctuations are independent of $\phi$ because our definition of $p_{ti}$ disregards direction.  

Equations  (\ref{eq:MultAns}) and (\ref{eq:corrFun0}) have two striking implications when combined with flow results from the previous section. First, if the variation of the initial geometric shape of the collision volume is the only source of fluctuations then (\ref{eq:uncorrRho}) implies that ${\cal R}  $ and $\langle \delta p_{t1}\delta p_{t2}\rangle$ must both vanish. Experiments have measured these quantities and they are both nonzero, as shown in Sec.\ \ref{glasma}. Second, the amplitudes $\overline{r}$ and $ \overline{\rho}_2$ determine the overall magnitude of correlations, as we see from (\ref{eq:flow2}). 
Anisotropic flow and momentum conservation determine the coefficients (\ref{eq:corrA}) and, therefore, the relative height of the near-side ``ridge'' at $\Delta\phi = 0$ and away-side features near $\Delta\phi = \pi$. 
However, interpretation of the evolution of the ridge height with beam energy or centrality requires an understanding of $\overline{r}$ or $\cal R$.

%
%	SECTION: Source of  Fluctuations 
%
\section{\label{source} Source of Fluctuations}

Nuclear collisions vary sharply from event to event due to differences in the number and configuration of the nucleons struck in the initial impact.  Each strike adds to a transient color field that lasts a proper time of roughly $\tau_0\sim 1$~fm. This field comprises an array of flux tubes connecting the fragments of the highly Lorentz-contracted nuclei along the beam direction. The number of participants determines the color charge and thus the overall strength of the fields. The flux tubes fragment after $\tau_0$, driving soft particle production. We emphasize that flux tubes arise naturally in QCD and have long been the core of phenomenological models such as PYTHIA.  In the next section we will focus on the Color Glass Condensate description, which incorporates these features in the high density environment produced by nuclear collisions and allows for systematic computations. For now, we keep the discussion more general.

We assume that the number and geometrical distribution of flux tubes is the most important source of fluctuations. Their fragmentation leads to density correlations described by
\be
c(\mathbf{x}_1, \mathbf{x}_2) = n_2(\mathbf{x}_1,\mathbf{x}_2) - n_1(\mathbf{x}_1)
n_1(\mathbf{x}_2),
\label{eq:CorrDef}
\ee
where $n_1$ and $n_2$ are the single and pair densities.  In the absence of correlations, $n_2(\mathbf{x}_1,\mathbf{x}_2) \rightarrow n_1(\mathbf{x}_1) n_1(\mathbf{x}_2)$ so that $c$ vanishes. The integral of $n_2$ over position gives the number of pairs averaged over events, $\langle N(N-1)\rangle$, so that  the integral of $c$ is ${\cal{R}  }\langle N\rangle^2$, with $\cal R$ given by (\ref{eq:MultFluctExp}).   We now take the flux tubes to be longitudinally boost invariant, so that the correlation function only depends on transverse coordinates  $\mathbf{r}_t = \mathbf{r}_{1,\, t} - \mathbf{r}_{2,\, t}$ as well as the average  $\mathbf{R}_t = (\mathbf{r}_{1,\, t} + \mathbf{r}_{2,\, t})/2$. 
We further assume particles from the same flux tube are primordially correlated and that correlations with the reaction plane arise due to the distribution of tubes. Since the transverse size of the flux tube is small, the primordial correlations reflect common spatial origins.
We then write  
\be
c(\mathbf{x}_1, \mathbf{x}_2)
 = \R{\la N\ra}^2\,\delta(\mathbf{r}_t ) \rho_{{}_{FT}} (\mathbf{R}_t).
\label{eq:CorrFunc}
\ee
Here, $\rho_{{}_{FT}}(\mathbf{R}_t)$ is the probability distribution for finding a flux tube at a transverse position $\mathbf{R}_t$ in the 
collision volume. This function describes the distribution of shape fluctuations discussed earlier.  Following Ref.\ \cite{Gavin:2008ev,Moschelli:2009tg}, we take $\rho_{{}_{FT}}$ to roughly follow the participant distribution of the colliding nuclei
\be
\rho_{{}_{FT}}(\mathbf{R}_t) \approx  \frac{2}{\pi R^2_A}  \lp 1 -\frac{R_t^2}{R_A^2}\rp
\label{eq:TubeDis}
\ee         
for $R_t \leq R_A$, and zero otherwise. 

We now discuss the impact of these long range correlations on the final-state distributions. Comoving partons locally thermalize as the flux tubes fragment.  Pressure builds and transverse expansion begins. In this process, partons from a flux tube at an initial space-time point $(t, \mathbf{x})$ will eventually acquire a final flow four velocity $u^\mu$. In a blast wave model  $u^\mu$ is taken to have a Hubble-like correlation, while hydrodynamic calculations provide a more realistic $u^\mu$. In either case, the Cooper-Frye single particle distribution is 
\begin{equation}
 \rho_1(\mathbf{p}) 
= \int f(\mathbf{x},\mathbf{p}) \, d\Gamma,
\label{eq:singles}
\end{equation}
where $f$ is one-body phase space distribution function and $d\Gamma = p^\mu d\sigma_{\mu}$ is the element of flux through the four dimensional freeze out surface.   In local equilibrium the distribution has the Boltzmann form $f(\mathbf{x},\mathbf{p}) =  {(2\pi)^{-3}}\exp\{-p^\mu u_\mu/T\}$.  The  temperature $T$ and fluid four-velocity $u_\mu$ are generally fixed by hydrodynamics, which enforces the local conservation laws.  In keeping with the boost-invariant distribution (\ref{eq:CorrFunc}), we assume that freeze out occurs at a proper time $\tau_F$, so that $p^\mu d\sigma_\mu = \tau_F m_t \cosh(y-\eta)d\eta d^2 r_t$, where $\eta = (1/2)\ln((t+z)/(t-z))$ is the spatial rapidity.

The pair distribution has an analogous Cooper-Frye form 
\begin{equation}\label{eq:rho2f2}
    \rho_2(\mathbf{p}_1,\mathbf{p}_2) = \int 
    f_2(\mathbf{x}_1,\mathbf{p}_1,\mathbf{x}_2,\mathbf{p}_2) d\Gamma_1d\Gamma_2,
\end{equation}
where $f_2$ is the two-particle Boltzmann distribution function. In local thermal equilibrium the two particle distribution is
\begin{equation}\label{eq:f2}
    f_2 = n_2(\mathbf{x_1},\mathbf{x_2})
    {{f(\mathbf{x_1},\mathbf{p_1})}\over {n(\mathbf{x}_1)}}
    {{f(\mathbf{x_2},\mathbf{p_2})}\over {n(\mathbf{x}_2)}},
\end{equation}
where $n_1$ and $n_2$ are the single-particle and pair densities discussed earlier, with $n(\mathbf{x}) = \int f(\mathbf{x},\mathbf{p}) d^3p$. This expression satisfies a generalization of the Boltzmann transport equation for $f_2$; the factors of $f$ cause the generalized collision terms to vanish just as they do in the one body equation. We omit momentum and energy conservation terms that do not contribute to $\cal{R}  $ and $\langle \delta p_{t1}\delta p_{t2}\rangle$.  

We understand (\ref{eq:f2}) as follows. In local equilibrium we can divide the system into fluid cells, each of which  is in equilibrium at the local temperature $T(\mathbf{x})$ and mean velocity $u^\mu(\mathbf{x})$. The momentum distribution in each cell must therefore be $f(\mathbf{x},\mathbf{p})$. The local equilibrium phase space distribution (\ref{eq:f2}) is correlated if there are density correlations between cells or autocorrelations. These correlations are described by the pair density $n_2(\mathbf{x_1},\mathbf{x_2})$.  
The integral $n_2$ over both positions gives the number of pairs averaged over events $\langle N(N-1)\rangle$.  In the absence of correlations, $ n_2(\mathbf{x}_1,\mathbf{x}_2) \rightarrow n_1(\mathbf{x}_1) n_1(\mathbf{x}_2)$. 

In order to study the angular distribution of fluctuations, we use (\ref{eq:singles}), (\ref{eq:rho2f2}), (\ref{eq:f2}),  and (\ref{eq:CorrDef}) to write (\ref{eq:corrFunExp}) as  
\be
r(\mathbf{p}_1, \mathbf{p}_2) =
\!\!\int c(\mathbf{x}_1, \mathbf{x}_2) 
f(\mathbf{x}_1,\mathbf{p}_1)
f(\mathbf{x}_2,\mathbf{p}_2)
d\Gamma_1d\Gamma_2.
\label{eq:MomCorr}
\ee
We consider the angular correlation function  
\begin{equation}\label{eq:deltaRho}
\Delta\rho = \!\!\int r \,\delta (\Delta\phi - \phi_1 + \phi_2)\delta(\Delta\eta - \eta_1 + \eta_2)d\mathbf{p}_{1}d\mathbf{p}_{2},
\end{equation}
where $r(\mathbf{p}_{1},\mathbf{p}_{2})$ is given by (\ref{eq:corrFunExp}). This function probes the $(\Delta\eta, \Delta\phi)$ correlations of particles in the full range of  $|\mathbf{p}_t|$.  Such correlations are dominated by the more abundant low $p_t$ particles. 

We follow Ref.\ \cite{Gavin:2008ev,Moschelli:2009tg} and identify $c(\mathbf{x}_1, \mathbf{x}_2)$ with (\ref{eq:CorrFunc}), a form that describes the system at its formation. This identification omits the effects of diffusion described in Ref.~\cite{Gavin:2006xd}. This omission is reasonable only as long as correlations are dominated by  pairs separated by  $|\eta_1 - \eta_2| > 1$. 

To clarify the contributions of fluctuations and anisotropic flow to correlations,  we expand (\ref{eq:deltaRho}) as a Fourier series. Equation (\ref{eq:DynamicMultDensity}) implies that the integral of  $\Delta\rho$ over $\Delta \phi$ in a rapidity range gives $\langle N\rangle^2{\cal{R}  }$. We therefore write
\begin{equation}\label{eq:fourier}
     {{\Delta\rho}\over{\langle N\rangle^2}}=  
        {{{\cal R}  }\over {2\pi}}\{1+2\sum_{n=1}^{\infty} A_n \cos(n\Delta\phi)\}. 
\end{equation}
Next, we compute the Fourier coefficients using (\ref{eq:MomCorr}) together with the correlation function (\ref{eq:CorrDef} -- \ref{eq:TubeDis}). We find
\begin{equation}\label{eq:floCo1}
A_n = \int d^2r_{t}  \rho_{{}_{FT}}(\mathbf{r}_t) [v_n(\mathbf{r}_{t})]^2
\end{equation}
where we define the local flow coefficient
\begin{equation}\label{eq:floCo2}
v_n(\mathbf{r}_t) = \int \!d\mathbf{p}\; {{f(\mathbf{r}_t,\mathbf{p})}\over{n(\mathbf{r}_{t1})}} \cos{n\phi}.
\end{equation}
%
%we have used of boost invariance to convert the $m_t \cosh(y-\eta)d\eta$ integrals in the Cooper-Frye expressions into a $dp_z$ integrals (the integrand is independent of $y$). 
Identifying the average over the probability distribution $\rho_{{}_{FT}}$  as the event average, we see that  (\ref{eq:floCo1}) has the form $A_n=\langle v_n^2 \rangle$ in (\ref{eq:corrA}); see Sec.\ \ref{Fluc}.  
An important caveat is that momentum conservation corrections omitted here will modify the low orders, particularly $A_1$ \cite{Borghini:2002mv,Borghini:2003ur,Borghini:2006yk,Borghini:2007ku,Chajecki:2008vg}.  We will treat these corrections and present a detailed computation of flow parameters and flow fluctuations elsewhere.   

We now compute $\langle\delta p_{t1}\delta p_{t2}\rangle$,  combining (\ref{eq:MomCorr}) with (\ref{eq:CorrDef} -- \ref{eq:TubeDis}) as before to evaluate (\ref{eq:Dynamic}). To simplify the denominator in (\ref{eq:Dynamic}), we note that (\ref{eq:MultFluctExp})
implies that $\langle N(N-1)\rangle = \langle N\rangle^2(1+{\cal R})$. We then obtain
\begin{equation}\label{eq:dpt1}
    \langle \delta p_{t1}\delta p_{t2}\rangle
    = {{{\cal R}}\over{1+{\cal R}}} \int d^2r_{t}  \rho_{{}_{FT}}(\mathbf{r}_t)[g(\mathbf{r}_{t})]^2,
\end{equation}
where the local momentum excess is
\begin{equation}\label{eq:Gdef2}
g(\mathbf{r}_t) = \int \!d\mathbf{p}\; {{f(\mathbf{r}_t,\mathbf{p})}\over{n(\mathbf{r}_{t})}} \left(p_t - \langle p_t\rangle\right).
\end{equation}
In contrast to the flow coefficients (\ref{eq:floCo1}), we see that $\langle\delta p_{t1}\delta p_{t2}\rangle$ is proportional to $\cal R$. However, both flow and fluctuation quantities depend on $\rho_{{}_{FT}}$ in a similar manner. Observe that $g$ vanishes if the velocity and temperature are uniform.

%
%	SECTION: Source of  Fluctuations 
%
\section{\label{glasma} Glasma Fluctuations}

The azimuthal dependence of the correlation function given by (\ref{eq:fourier}), (\ref{eq:floCo1}), and (\ref{eq:floCo2}) is essentially equivalent to that used in hydrodynamic models  \cite{Teaney:2010vd}.  The relative height of the near-side ridge compared to the away-side features in $\Delta\phi$ depends on the shape fluctuations $\rho_{{}_{FT}}$, the magnitude of the flow coefficients, and momentum conservation.  
However, the overall scale of correlations is set by the multiplicity fluctuations $\cal{R}$. This scale is crucial if one wishes to compare correlations at different centralities or beam energies. The difficulty with interpreting the shape of the correlation function is that one must disentangle information on the production mechanism contained in $\rho_{{}_{FT}}$ from flow and viscosity effects. Unless a number of simplifying assumptions hold true, this may prove challenging  \cite{Bhalerao:2011yg}.   

%We argue in this section that $\cal{R}$ can be measured directly using (\ref{eq:MultFluctExp}). This sidesteps the complication of flow, as discussed in Sec. \ref{Fluc}. 
%
It is easiest to appreciate the significance of $\cal{R}$ in the context of Color Glass Condensate theory. 
Suppose that each collision produces $K$ flux tubes, and that this number varies from event to event with average $\langle K\rangle$. Each Glasma flux tube yields an average multiplicity of  $\sim \alpha_s^{-1}(Q_s)$ gluons, where $Q_s$ is the saturation scale \cite{Kharzeev:2000ph}.  The number of gluons in a rapidity interval $\Delta y$ is then
%
%	dN/dy
\be
\langle N\rangle = ({{dN}/{dy}})\Delta y  \sim {\alpha_s}^{-1}(Q_s)\langle K\rangle.
\label{eq:Nscale}
\ee
In the saturation regime $K$ is proportional to the transverse area $R_A^2$ divided by the area per flux tube, $Q_s^{-2}$ \cite{Kharzeev:2000ph}. 
In Ref.~\cite{Gavin:2008ev} we show that the scale of correlations is set by 
%
%	R definition
\be
{\cal R} = {{\langle N^2\rangle - \langle N\rangle^2 - \langle N\rangle}\over{\langle N\rangle^2}} \propto \langle K\rangle^{-1}.
\label{eq:Rdef}
\ee
The dependence on $\langle K\rangle$ drops out of the product 
%
%	RdN/dy
\be
{\cal R}{{dN}/{dy}} = \kappa {\alpha_s^{-1}}(Q_s),
\label{eq:CGCscale}
\ee
a result consistent with calculations of Dumitru et al. in Ref.~\cite{Dumitru:2008wn}.  Gelis, Lappi, and McLerran have shown that the multiplicity distribution $P(N)$ in the Glasma follows a negative binomial distribution \cite{Gelis:2009wh}. We point out that by definition their negative binomial parameter $k_{{}_{NBD}}$ satisfies ${\cal R}= 1/k_{{}_{NBD}}$. Their calculated $k_{{}_{NBD}}$ agrees with (\ref{eq:Rdef}) and (\ref{eq:CGCscale}). 

The signature of the Glasma contribution to correlations and fluctuations is that ${\cal R}dN/dy$ depends only on $Q_s$. Equation (\ref{eq:CGCscale}) therefore constitutes a scaling relation, since $Q_s$ depends on many collision variables in a combination that can be computed from first principles. The leading order formula in Ref.\ \cite{Kharzeev:2000ph}  relates $Q_s$ to the beam energy and the number of participants per unit area, which in turn depends on $A$ and $b$. Measurements of the ridge at various beam energies, target masses, and centralities fix the dimensionless coefficient $\kappa$ in (\ref{eq:CGCscale}) and are in excellent accord with the leading-order dependence  \cite{Gavin:2008ev,Moschelli:2009tg}.  Uncertainties in the underlying description of flow were the biggest source of uncertainty in comparing  (\ref{eq:CGCscale}) to ridge data. 

\begin{figure}
%\vskip -0.5in
\centerline{\includegraphics[width=3.4in]{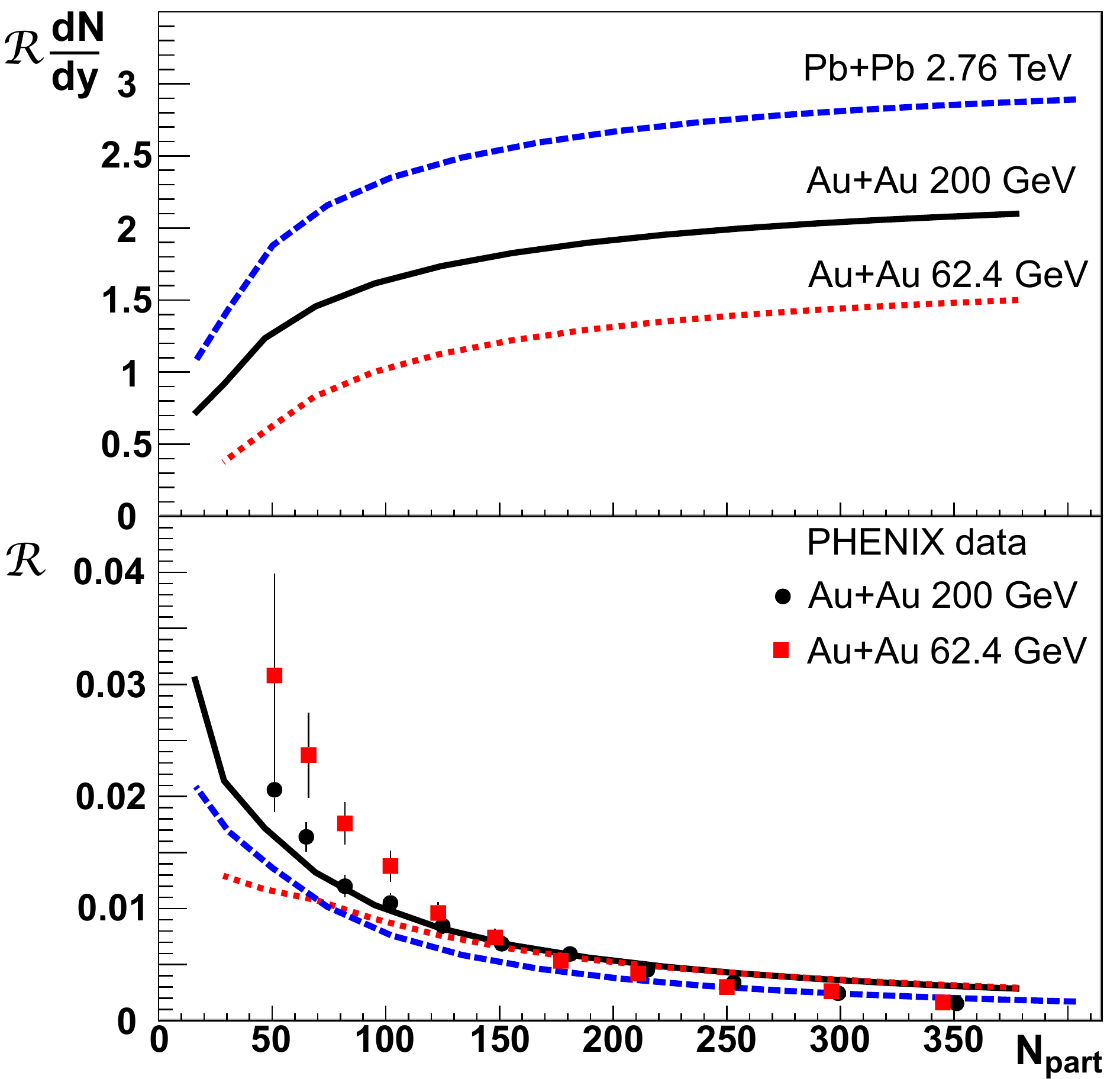}}
%\centerline{\includegraphics[width=3.4in]{RdNdy_Npart.pdf}}
%\vskip -0.4in
\caption[]{Prediction for ${\cal R}dN/dy$ as a function on the number of participants $N_{\rm part}$ at three beam energies (top). Calculated $\cal R$ compared to PHENIX data from \cite{Adare:2008ns} (bottom).}
\label{fig:fig1}\end{figure}

Multiplicity fluctuation measurements of $\cal R$ in principle circumvent the complexity of flow, facilitating the search for this Glasma scaling or other production mechanism signatures.  As discussed in Sec.\ \ref{Fluc},  $\cal{R} $ integrates the correlation function (\ref{eq:MomCorr}), so that the $\cos(n\Delta\phi)$ contributions to (\ref{eq:fourier}) vanish. Predictions shown in Fig.\ \ref{fig:fig1} are obtained using (\ref{eq:CGCscale}) with the energy independent dimensionless coefficient $\kappa$ fixed to fit the near side ridge as in \cite{Gavin:2008ev,Moschelli:2009tg}. The number of participants $N_{\rm part}$ is used to indicate centrality.  

PHENIX has measured multiplicity fluctuations at RHIC \cite{Adler:2007fj,Adare:2008ns}. They report a negative binomial parameter $k_{{}_{NBD}}$. The best we can do is to compare their $k_{{}_{NBD}}^{-1}={\cal R}$ to  (\ref{eq:CGCscale}) divided by $dN/dy$ measured elsewhere. Results shown in the lower panel in Fig.\ \ref{fig:fig1} agree fairly well in central collisions. 

The experimental result that $\cal R$ is not zero shows that geometric shape variation is not the only source of fluctuations. To deduce anything beyond that, two caveats are in order. First, (\ref{eq:Rdef}) and (\ref{eq:CGCscale}) strictly apply only to the number of gluons. Taking the number of particles to be conserved through hadronization as discussed in Ref.\ \cite{Kharzeev:2000ph},  we can identify $\cal R$ in (\ref{eq:Rdef}) with the measured multiplicity fluctuations (\ref{eq:MultFluctExp}). However,  one is then unsure how to address phenomenological concerns such as resonance decay. Second, experimenters must exercise care in measuring $\cal{R}$ as a function of centrality, because centrality selection can distort multiplicity fluctuations.  Using narrow multiplicity bins to select centrality will remove fluctuations entirely.  One way to remove this bias is to use a zero degree calorimeter to select centrality as PHENIX has done \cite{Adare:2008ns}. See the appendix for details. 
%
%\cite{Kafka:1975cz,Thome:1977ky,Alner:1985zc,Alner:1987wb,Arneodo:1987qy,Adamus:1987ea,Ansorge:1988kn,Abbott:1995as,Bachler:1992jp,Albrecht:1989kh,Aggarwal:2001aa,Aggarwal:2001aa,Alt:2006jr} 
%}
%

An alternative probe of Glasma scaling is $\langle \delta p_{t1}\delta p_{t2}\rangle$. As with $\cal R$, the anisotropic flow contributions vanish on angular integration. The quantity $\langle \delta p_{t1}\delta p_{t2}\rangle$ is designed to be independent of multiplicity fluctuations, reducing our hadronization concerns \cite{Voloshin:1999yf,Voloshin:2001ei}. Moreover, it is effectively free of the multiplicity bias effect, as shown in the appendix.  Unlike $\cal R$, momentum fluctuations depend on the scale of the fluid velocity because flow enhances $p_t$.  We can constrain this dependence using $v_2$ and $\langle p_t\rangle$ measurements. 

In order to calculate $\langle \delta p_{t1}\delta p_{t2}\rangle$ from (\ref{eq:dpt1}) and (\ref{eq:Gdef2}) we must specify the relation between the initial transverse position $\mathbf{r}_t$ of a fluid cell and its final transverse flow velocity $\mathbf{v}_t$.  To maintain consistency with our ridge analysis in Ref.\ \cite{Gavin:2008ev,Moschelli:2009tg}, we use a blast wave model.  
\begin{figure}
%\vskip -0.5in
\centerline{\includegraphics[width=3.3in]{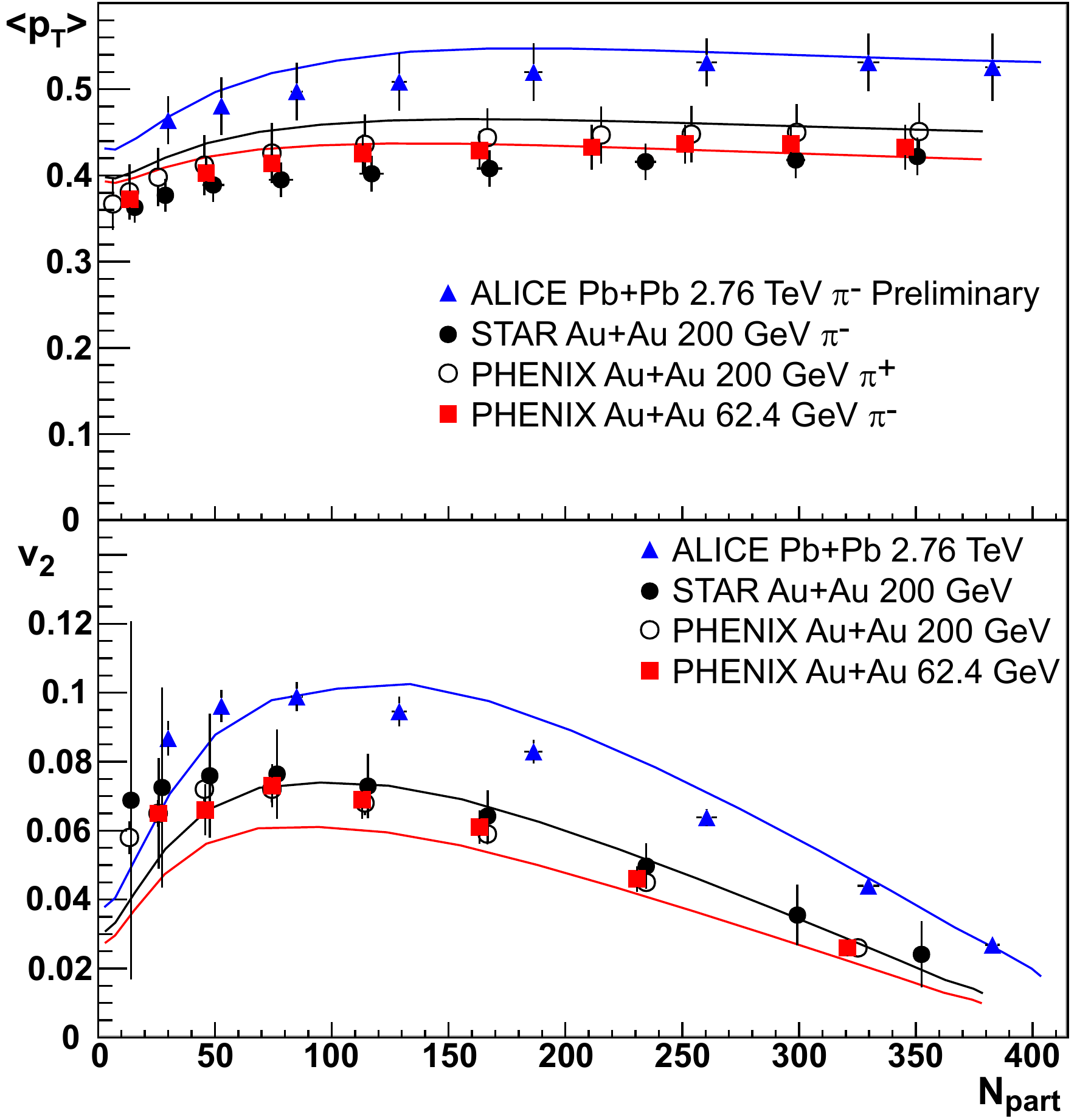}}
%\vskip -0.4in
\caption[]{Average transverse momentum (top) and elliptic flow $v_2$ (bottom) as functions of $N_{\rm part}$ for the same energies as in Fig.\ \ref{fig:fig1}. Mean $p_t$ data in GeV are from \cite{Adler:2003cb,Abelev:2008ez,Collaboration:2011rt} and $v_2$ data are from \cite{Shimomura:2009zz,Adams:2004bi,Aamodt:2011vk}. }
\label{fig:fig2}\end{figure}
There we assumed $\gamma \mathbf{v}_t = \lambda \mathbf{r}_t$ and uniform temperature $T$, with parameters chosen to reproduce fits to 200 GeV Au+Au $p_t$ spectra \cite{Gavin:2008ev,Moschelli:2009tg}.  Sorensen et al.\  subsequently pointed out the importance of allowing for the ellipticity of the source in describing the centrality dependence of the ridge \cite{Sorensen:2011hm}.  To account for the ellipticity of the collision volume, we now take $\gamma \mathbf{v}_t = \lambda (\epsilon_x x \hat{x} + \epsilon_y y \hat{y})$, where $\epsilon_{x,y}^2 = 1\pm \epsilon$. The average $v_t$ and freeze out temperature in 62 GeV and 200 GeV Au+Au collisions are the same as those used in \cite{Gavin:2008ev,Moschelli:2009tg} and are based on an analysis in \cite{Kiyomichi:2005zz}.  At 2.76 TeV the velocity is scaled up from the 200 GeV values by 6\% and the temperature is scaled up by 7\%. We present these parameters in Fig.\ \ref{fig:fig2.5}; note that the change in $v_t$ and $T$ with beam energy is rather small. The eccentricity $\epsilon$ is chosen to fit the observed centrality dependence of elliptic flow $v_2$ in 62 and 200~GeV Au+Au collisions and 2.76 TeV Pb+Pb collisions \cite{Shimomura:2009zz,Adams:2004bi,Aamodt:2011vk}. 
The top panel in Fig.\ \ref{fig:fig2} compares these calculations to data from Refs.\ \cite{Adler:2003cb,Abelev:2008ez,Collaboration:2011rt}. We then use (\ref{eq:floCo2}) to compute $v_2$ and compare to $v_2\{2\}$, which includes fluctuations. 
To see how well these blast wave parameterizations work for the problem at hand, we calculate the average transverse momentum $\langle p_t\rangle = \int p_t \rho_1d\mathbf{p} /\int \rho_1d\mathbf{p}$ using  (\ref{eq:singles}). The agreement is shown in Fig.\ \ref{fig:fig2}.

\begin{figure}
%\vskip -0.5in
\centerline{\includegraphics[width=3.25in]{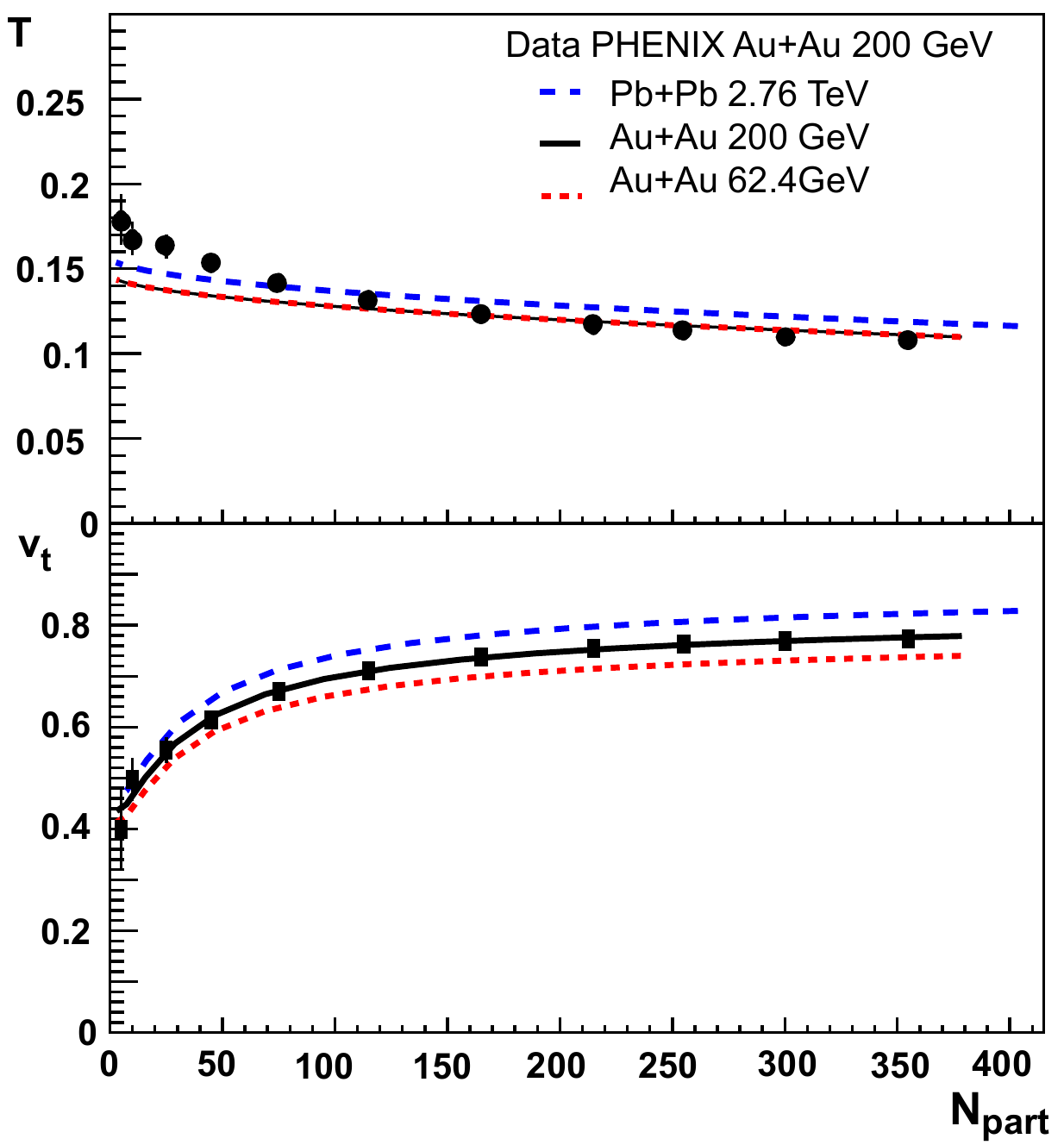}}
%\centerline{\includegraphics[width=3.2in]{RdNdy_Npart.pdf}}
%\vskip -0.4in
\caption[]{Blast-wave parameters for temperature (in GeV) and average velocity as functions of centrality. Data are from Ref.\ \cite{Kiyomichi:2005zz}.}
\label{fig:fig2.5}\end{figure}
\begin{figure}
%\vskip -0.5in
\centerline{\includegraphics[width=3.5in]{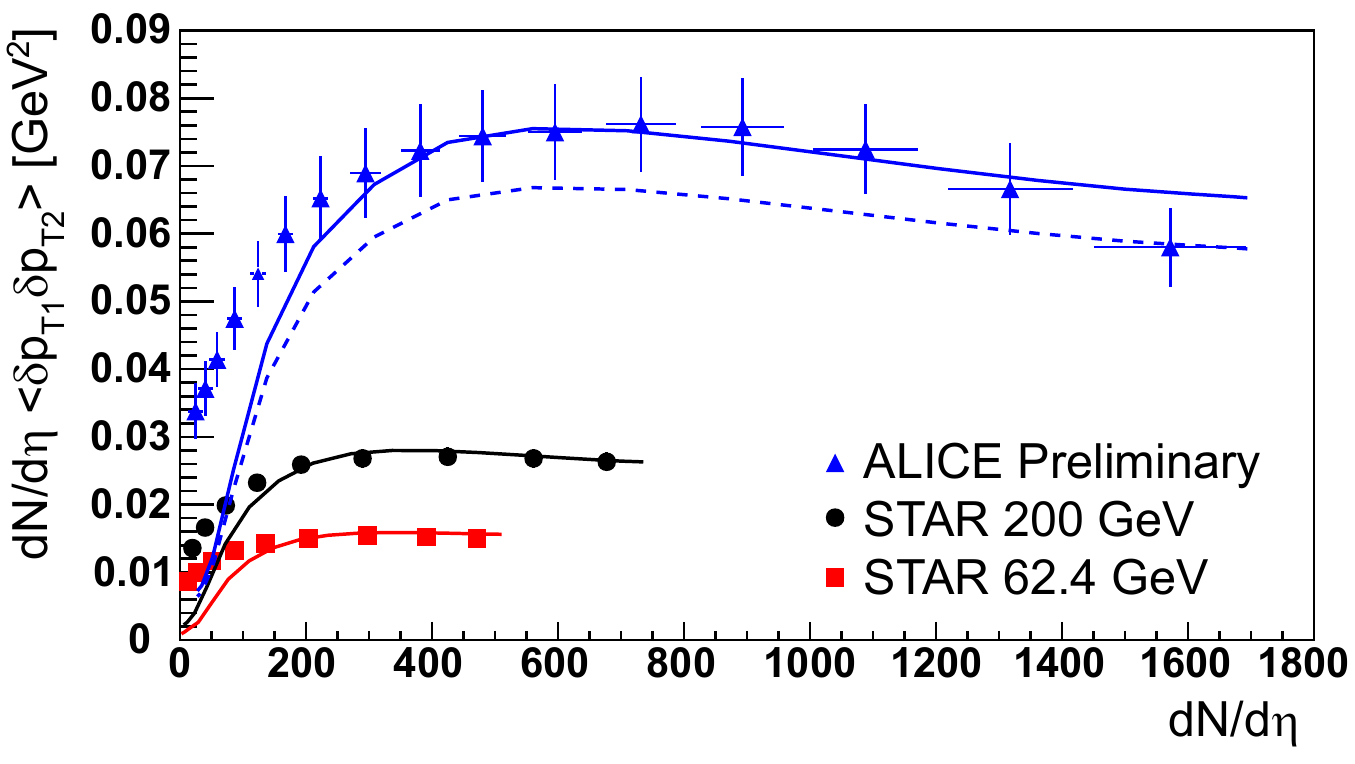}}
%\vskip -0.4in
\caption[]{ Transverse momentum fluctuations $\langle \delta p_{t1}\delta p_{t2}\rangle dN/d\eta$ as a function on the number of participants $N_{\rm part}$ at the same three beam energies. Data is from  \cite{Adams:2005ka,Heckel_ALICEptFluc}. Dashed and solid 2.76 TeV curves represent different $K$ and $p$ contributions to charged particle fluctuations as discussed in the text.}
\label{fig:fig3}\end{figure}

We now compute $\langle \delta p_{t1}\delta p_{t2}\rangle$. The multiplicity variance $\cal R$ is obtained from (\ref{eq:CGCscale}).  We employ (\ref{eq:dpt1}) and (\ref{eq:Gdef2}) combined with (\ref{eq:TubeDis}), and use the blast wave parameters discussed above. These are compared to data from Ref.\ \cite{Adams:2005ka,Heckel_ALICEptFluc} in Fig.\ \ref{fig:fig3}. The STAR collaboration at RHIC and the ALICE collaboration at LHC measure $\langle \delta p_{t1}\delta p_{t2}\rangle$ for charged particles rather than pions. There is little difference between these quantities at RHIC, but at LHC the $K/\pi$ and $p/\pi$ ratios are appreciably larger than expected in the observed range $0.15 < p_t < 2$~ GeV \cite{Floris}.  Using the measured $\langle p_t\rangle$ and particle ratios  for kaons and protons gives $\langle p_t\rangle_{ch}/ \langle p_t\rangle_{\pi}\sim 1.13$; PYTHIA gives $\sim 1.07$. The top solid curve in Fig.\ \ref{fig:fig3} is our computation with the measured $K/\pi$ and $p/\pi$ ratios, while the dashed curve assumes PYTHIA ratios. 
Agreement with data is very good for central collisions where our local equilibrium assumptions are most applicable. Deviations in peripheral collisions may be due in part to incomplete thermalization, see \cite{Gavin:2003cb,Gavin:2004dc}.    
%

%
%	SECTION:	SUMMARY
%
\section{\label{summary}Summary}

In this paper we have studied the connection between long range correlations, fluctuations, and flow. While our results in Sec.\ \ref{glasma}  primarily address fluctuations, we discuss flow in Sec.\  \ref{source} to establish its roll in correlations so that we can isolate its effect. As anticipated, our flow coefficients and their fluctuations given by (\ref{eq:floCo1}) and (\ref{eq:floCo2}) depend primarily on the spatial distribution of flux tubes -- the ``shape fluctuations'' -- described by the probability distribution $\rho_{{}_{FT}}$.  Consequently, the relative magnitude of near and away-side features in $\Delta\phi$ correlations depends on these shape fluctuations. However, it is only in this relative sense that hydrodynamic response to initial shape fluctuations  explains long range correlations \cite{Luzum:2011mm}.  

The scale of correlations $\cal{R}$ is important in comparing correlations at different centralities and energies. In particular, Refs.\ \cite{Gavin:2008ev,Moschelli:2009tg} compared calculations to the peak height of the ridge at $\Delta\phi=0$ for Au+Au and Cu+Cu collisions for a range of energies and kinematic conditions. Experimenters normalize the peak $\Delta\rho$ to the number of pairs from mixed events $\sqrt{\rho_{ref}}$. When jet and momentum conservation contributions are omitted, our results had the form $\Delta\rho/\sqrt{\rho_{ref}} = {\cal R} (dN/dy) F(\Delta\phi)$, where $F$ depends only on flow. We found the beam-energy, centrality,  and $A$ dependence of ${\cal R}$ to be consistent with data; see \cite{Moschelli:2009tg} for details. To relate this to the current context, observe that the Fourier decomposition of $F$ gives (\ref{eq:fourier}). This supports the arguments regarding the physical significance of the height of the ridge stated in \cite{Dumitru:2008wn,Gavin:2008ev,Dusling:2009ni,Moschelli:2009tg}.  A  limitation of Refs.\  \cite{Dumitru:2008wn,Gavin:2008ev,Dusling:2009ni,Moschelli:2009tg} is that it is difficult to distinguish flow from Glasma effects when concentrating exclusively on the ridge.  

Fluctuation studies provide an alternative set of experimental techniques for attacking correlation physics. In Sec.\ \ref{Fluc} we showed that multiplicity and $p_t$ fluctuations  are independent of anisotropic flow. Measurement of such fluctuations remove much of the hydrodynamic uncertainty of ridge studies \cite{Dumitru:2008wn,Gavin:2008ev,Dusling:2009ni,Moschelli:2009tg}.  
We emphasize that the mere fact that the measured $\cal R$ and $\langle \delta p_{t1}\delta p_{t2}\rangle$ are nonzero proves that event-wise variation of the initial geometric shape is not the only source of fluctuations.

In Sec.\ \ref{glasma} we related the correlation strength to two fluctuation observables $\cal R$ and $\langle \delta p_{t1}\delta p_{t2}\rangle$. 
%The  contribution of long range correlations to multiplicity fluctuations $\cal R$ are entirely independent of flow, but have other experimental difficulties. 
Glasma calculations of $\cal R$ in Fig.\ \ref{fig:fig1}  and  $\langle \delta p_{t1}\delta p_{t2}\rangle$ in Fig.\ \ref{fig:fig3} are in good accord with data, except for peripheral collisions. 
%
%Glasma calculations of $p_t$ fluctuations in Fig.\ \ref{fig:fig3} are again in excellent agreement with data.  
The deviation in peripheral collisions may reflect the breakdown of the assumption of local equilibrium. The onset of thermalization in peripheral collisions modifies the correlation function.  In particular, this effect has been shown to modify $\langle \delta p_{t1}\delta p_{t2}\rangle$ at low numbers of participants \cite{Gavin:2003cb}. Partial thermalization describes peripheral RHIC data very well and, moreover, allows one to describe $pp$ and $AA$ collisions in the same model \cite{Gavin:2004dc}. 

The fact that long range correlations can account for the measured $\langle \delta p_{t1}\delta p_{t2}\rangle$ leaves us with a puzzle. There are two types of fluctuations in nuclear collisions. First, each collision produces a different multi-particle system. Second, the partonic system produced in each event undergoes dynamic fluctuations as it evolves and hadronizes.  We have focused on the source of long range correlations, where causality precludes the second type. However, fluctuation observables integrate both long and short range effects. Short range effects include jets and jet quenching, HBT, resonance decay, hadronization and hadro-chemical effects -- not to mention novel fluctuations due to the phase transition. These effects modify correlations in the range $\Delta\eta \sim 1 -2$ units or less.  Jet-quenching, which most strongly affects central collisions, was estimated in Ref.~\cite{AbdelAziz:2005ds}. Long range effects such as flow are charge independent. Net-charge fluctuations are therefore sensitive primarily to short range effects and, indeed, data show the appropriate rapidity dependence \cite{Abelev:2008jg}.  Perhaps measurement of charge-dependent fluctuations, ${\cal R}_{ab}$ and $\langle \delta p_{t1}\delta p_{t2}\rangle_{ab}$ for $a,b=+$ or $-$, would help distinguish long range from short range contributions.

%We explore the proposition that only the event-by-event fluctuations matter.  

%
%	SECTION:	ACKNOWLEDGMENTS
%
\begin{acknowledgments}
S.G.\ thanks C.\ Pruneau, P.\ Steinberg, L.\ Tarini, R.\ Venugopalan, S.\ Voloshin, and C.\ Zin for discussions. G.M.\ thanks M.\ Bleicher and S. Heckel for discussions and support. This work was supported in part by U.S. NSF grants PHY-0855369 (SG) and The Alliance Program of the Helmholtz Association (HA216/EMMI) (GM).
\end{acknowledgments}

%
%	SECTION:APPENDIX
%
\appendix
\section{\label{appendix}Experimental Considerations}

In this appendix we discuss experimental issues that affect the measurement of the fluctuation observables ${\cal R}$ and $\langle \delta p_{t1}\delta p_{t2}\rangle$. We show that biases can modify ${\cal R}$ if multiplicity is used to identify centrality in fluctuation observables, but that $\langle \delta p_{t1}\delta p_{t2}\rangle$ is largely unaffected. Experimental considerations in fluctuation measurements were discussed in Ref.\ \cite{Pruneau:2002yf}. These observables were constructed to be independent of experimental acceptance and efficiency effects, and many contributions that alter $\cal R$ were estimated. We extend the treatment in Ref.~ \cite{Pruneau:2002yf} to include $\langle \delta p_{t1}\delta p_{t2}\rangle$. 

Assume that the collision produces $K$ independent `sources' which then produce particles; one can think of each source as a flux tube or a wounded nucleon, depending on one's favorite model. A set of events with fixed $K$ produces the single and pair densities that scale as:
\begin{equation}\label{eq:sources}
\rho_1 = \hat{\rho}_1 K, \,\,\,\,\,\,\,\, \rho_2 = \hat{\rho}_2 K + \hat{\rho}_1 \hat{\rho}_1 K (K-1).
\end{equation}
Suppose that experimenters measure a multiplicity $m$ to identify centrality. Each $m$ bin receives contributions from a range of $K$, so that the measured quantities are averaged over that ensemble. 
The multiplicity of a particular particle species will roughly satisfy 
$\langle N\rangle_m = \langle\int \rho_1\rangle_m \propto  \langle K \rangle_m$. 
%%
%\begin{equation}
%\langle N\rangle_m = \langle\int \rho_1\rangle_m= \mu \langle K \rangle_m, 
%\end{equation}
%%
%where $\mu \equiv \int \hat{\rho}_1$. 
Equations (\ref{eq:MultFluctExp}), (\ref{eq:DynamicMultDensity}), and (\ref{eq:sources}) imply that multiplicity fluctuations satisfy 
\begin{equation}\label{eq:RscaleMult}
 {\cal R}_m =  {{A}\over{\langle K\rangle_m}} + {{\langle K^2\rangle_m -\langle K\rangle_m^2}\over{\langle K\rangle_m^2}}. 
\end{equation}
%
% %
%\begin{equation}\label{eq:RscaleMult}
% {\cal R}_m = 
%{{\sigma_n- \mu^2}\over{\mu^2}} {{1}\over{\langle K\rangle_m}} + {{\langle K^2\rangle_m -\langle K\rangle_m^2}\over{\langle K\rangle_m^2}}. 
%\end{equation}
%%
%for $\sigma_n \equiv \int \hat{\rho}_2$. 
where $A$ is a model-dependent constant, see Ref.~ \cite{Pruneau:2002yf} for details. The first term represents the average fluctuations per source. The second term comes from fluctuations in the number of sources. Unconstrained, independent sources follow Poisson statistics, $\langle K^2\rangle_m -\langle K\rangle_m^2 = \langle K\rangle_m$.  

To illustrate the effect of centrality cuts on ${\cal R}_m$, consider the following opposite extremes. Suppose for the moment that $m$ and $N$ are multiplicities measured in the same rapidity interval, so that one can approximate $m \propto K\propto N$. Then the fluctuations of $K$ for fixed $m$ vanish.  On the other hand, if we take $m$ to be the signal in a zero degree calorimeter, then $m$ and $K$ are correlated only by the impact parameter $b$. Fluctuations of $K$ may then dominate (\ref{eq:RscaleMult}) and be Poissonian \cite{Gavin:2008ev}.  

We now consider transverse momentum fluctuations $\langle \delta p_{t1}\delta p_{t2}\rangle$. We combine (\ref{eq:sources}) with (\ref{eq:Dynamic}), first noting that $\langle p_t\rangle = \int p_t \rho_1/\int \rho_1$ implies that $\int \rho_1 \delta p_t \equiv 0$. Therefore, the numerator of  (\ref{eq:Dynamic}) satisfies  $\int \rho_2 \delta p_{t1} \delta p_{t2} \propto \langle K\rangle_m$. The denominator of (\ref{eq:Dynamic}) is $\langle N(N-1)\rangle_m = \langle N\rangle_m^2(1+{\cal R}_m)$. This implies that 
\begin{equation}\label{eq:DynamicMult}
   \langle \delta p_{t1}\delta p_{t2}\rangle_m =
    {{B}\over{\langle K\rangle_m}} \, {{1}\over{1+{\cal R}_m}},
\end{equation}
where $B$ is another constant. 
This quantity has no additional contribution from $K$ fluctuations as in (\ref{eq:RscaleMult}). However, a small centrality-bias effect may result from the ${\cal R}_m$ in the denominator. 

To estimate this effect, observe that our CGC calculation gives ${\cal R}\sim 0.0017$ for central 2.76 GeV PbPb collisions, and 0.02 for the most peripheral value computed, which corresponds to $N_{\rm part} \approx 10$ participants. The effect of centrality bias on $\langle \delta p_{t1}\delta p_{t2}\rangle$ is therefore negligible. However, it might be important in pp collisions with a multiplicity trigger. Of course, the effect could be eliminated by normalizing to $\langle N\rangle^2$ rather than $\langle N(N-1)\rangle$, but that would have consequences regarding the cancellation of efficiencies.

\bibliography{ridge_fluctuations_refs}

\end{document}